# Polarization independent hot electron production from modulated solid surfaces

P. P. Rajeev and G. Ravindra Kumar*

*Tata Institute of Fundamental Research,
1, Homi Bhabha Road, Colaba, Mumbai – 400 005*

## Abstract

We present measurements of hard x-rays in the 50-300 keV range from copper plasmas produced by 100 fs, 806 nm laser pulses at a peak intensity ~ $10^{16}$ W cm$^{-2}$. Surface roughness, even at the tens of nanometer level, is shown to influence the emission characteristics. The enhanced emission from rough targets is attributed to depolarization of light as well as extra absorption facilitated by the surface irregularities.

* Corresponding author .Tel 91 22 215 2971; fax 91 22 2152110

*E-mail address*: grk@tifr.res.in.

# Introduction

Intense, ultrashort lasers are being used extensively to explore light-matter interaction above the breakdown threshold of a solid [1]. The wide ranging interest in this subject is partly due to the promise of the resulting short-lived solid density plasmas as bright sources of extremely energetic material particles and ultrashort x-rays [2-4]. The potential for creating high brightness x-ray sources, in particular, has triggered research to explore various applications like x-ray lithography and time resolved x-ray diffraction [1]. The progress in understanding the fundamental processes in such plasmas is being accompanied by efforts for more efficient coupling of laser energy, so as to enhance the x-ray production. In addition to impressive brightness levels, an exciting property of such x-ray pulses [5] is their extremely short temporal duration (subpicosecond), which is ideal for time resolved studies at x-ray wavelengths. There is a great deal of interest in methods that could enhance the x-ray yield and the influence of various laser and target conditions has been the subject of many recent studies. Laser pre-pulses have been investigated in detail as yield enhancers. While significant enhancement in the yields is noticed, the x-ray pulse duration tends to become longer [6, 7]. In the recent past, there has been an increased interest in the role of modulation of the target surface in the efficient coupling of input light into the plasma with resultant increase in x-ray yields. Murnane and coworkers [8] have shown absorption of over 90% of incident light into the plasma formed on grating targets as well as those coated with metal clusters. More recently, impressive enhancements of x-ray flux have been achieved in nanohole alumina targets (soft x-ray region) [9], porous silicon [10] and nickel `velvet' targets (mildly hard x-ray region) [11]. Methods of enhancing emission in the very hard x-ray spectral region are hardly explored. It is important to understand the role of roughness in enhancing the production of hot electrons in the plasma, which in turn are responsible for the hard x-ray generation. During the course of our studies on x-ray emission, we observed that easily available, unpolished targets showed an enhancement in the yield in both hard and very hard x-ray regimes apart from removing the polarization dependence of the emission [12]. We have also shown recently that nanoparticle coated targets can be excellent emitters of very hard x-rays [13]. In this paper, we investigate the influence of surface roughness on hard x-ray generation. We demonstrate the positive effect of mild to moderate (sub-wavelength) surface roughness on x-ray yields in the 30- 300 keV region.

## 2. Experimental Details

A schematic of the experimental set up is shown in Fig.1. A Ti: Sapphire laser operated at 806 nm, 100 fs was focused with a 30 cm focal length lens on to copper targets housed in a vacuum chamber at $10^{-3}$ Torr. The femtosecond laser is a custom-built chirped pulse amplification system with two-stage amplification, which can generate 50 mJ, 100 fs pulses. The maximum pulse energy used in the current experiments is 6 mJ, giving a maximum focused intensity of about $10^{16}$ W cm$^{-2}$ with a focused spot size of 30 µm. The laser has a prepulse (13 ns ahead of the main pulse) that was at least $10^4$ times weaker and the contrast with the picosecond pedestal was better than $10^5$. Under these conditions, plasma formation by prepulse/pedestal is found to be negligible [14]. To prove the insignificance of a preplasma in our experiment, we present the measurements (Fig.1, Inset) of time resolved reflectivity from the plasma, which was obtained using standard pump-probe geometry. The rapid, near step like fall in the reflectivity of the weak probe at zero-delay time (corresponding to the probe being reflected by the plasma formed by the pump instead of the metal) is indicative of a steep plasma density gradient and hence, the lack of a preplasma shelf [15]. A thin half wave plate was introduced in the beam path in order to change the polarization states of the light. The target was constantly rotated and translated to avoid multiple hits at the same spot by the laser pulses. X-ray emission from the plasma was measured at 45° to the plasma plume, which is along the normal to the target, using a Na I (Tl) scintillation detector. The 3mm thick glass window of the vacuum chamber sets a lower energy cutoff of about 20 keV for the observed emission. The output of the detector was amplified and fed to a multichannel analyzer through an ADC. The detector was shielded by lead bricks and calibrated using $Co^{57}$, $Cs^{137}$ and $Eu^{152}$. The temperature fits are obtained using the data above 50 keV, where the transmission is 100%, so as to ensure their reliability. The count rate was reduced to less than 0.1-0.2 per laser shot by introducing suitable lead apertures in front of the detector, so as to minimize the probability for pile-up. Further, the detector was kept about 30-40 cm away from the chamber window to prevent detection of spurious emissions from the chamber walls. Nearly background free spectra were obtained by eliminating cosmic ray noise using time gating - the laser pulse trigger was sent to a delay gate generator, which activated a time window of 10-20 microseconds for the signal acquisition. Spectra were typically collected over thousands of laser shots.

## 3. Results and Discussions

Fig. 2 shows the variation of bremsstrahlung emission (50 – 125 keV) from an optically polished target (local roughness < 5nm, AFM image shown in the inset) with incident angle for p-polarized light irradiation, keeping the intensity constant (corrected for oblique incidence) at $10^{16}$ Wcm$^{-2}$. The emission peaks broadly around 35°, indicating the influence of Resonance absorption (RA) in the laser coupling process [16]. RA efficiency reduces drastically for near-normal as well as near-grazing angles. In the above figure, the x-ray yield does not vanish at small angles because inverse bremsstrahlung (IB), which is not so sensitive to the angle of incidence, also plays a role in the laser coupling to the plasma. RA has been well studied experimentally [17] and its dependence on the plasma scale length (L) and the angle of incidence has been well characterized. The absorption peak coincides with the Fresnel value for extremely steep density gradients and shifts to lower angles as the plasma scale length increases [18, 19].

In RA, the absorbed energy flux peaks at $\theta_{max} = \sin^{-1}[0.8(k_0 L)^{-1/3}]$, for a linear density profile [16]. Assuming that the hot electron production also peaks at the same point, we deduce an approximate density scale length $L = 0.4\lambda$ from the above data, which is consistent with the other measurements in femtosecond plasmas [17]. However, more accurate and detailed calculations of absorption [19], with a realistic 10% error in our angular measurements, suggest a scale length $\sim 0.1\ \lambda$. Even though these calculations are performed for ultraviolet light, the wavelength is incorporated in dimensionless quantities viz. $n_e/n_{cr}$ and $\nu/\omega$, ($n_e$ is the electron density, $n_{cr}$, the critical density, $\nu$, the collisional frequency and $\omega$, the laser frequency) with values similar to our case, for a different material and for a different wavelength. Though the conventional RA formula is applicable for plasmas with $L \geq \lambda$, Fedosejevs et al. [19] show that it agrees quite well with more involved calculations for $L > 0.1\ \lambda$. However, the scale length extracted from the above measurement is only approximate. Detailed interferometric measurements [20] are necessary to compute the electron density scale lengths accurately.

Fig. 3 provides the Bremsstrahlung emission in 50 – 300 keV range from a reasonably polished copper target with local surface roughness ~ 25 nm and an unpolished one with average local roughness > 1μm, irradiated with p-polarized light at 10°. The integrated x-ray emission per second, assuming isotropic emission, in the above range from the polished target is $6.5 \times 10^{-10}$ J where as it is $2.9 \times 10^{-9}$ J from its unpolished counterpart, which amounts to a 4-fold enhancement in the x-ray yield. The least-square exponential fits yield a hot electron temperature component of ~ 15 keV for both the targets and a higher component of ~ 50 keV in the case of the unpolished one. For the polished target counts with energy > 125 keV were too small to give a meaningful higher temperature component. The inset shows the enhancements obtained using a similar unpolished target (average local roughness > 1 μm) with respect to an optically polished copper piece (average local roughness < 5 nm), as a function of angle of incidence. Huge enhancements are observed at angles close to normal incidence, and they decay monotonically as the incident angle in increased.

In the case of an optically flat target, at angles close to normal incidence, only collisional absorption (inverse bremsstrahlung) is the dominant light coupling mechanism, irrespective of the light polarization [16]. However, any realistic focusing geometry could change the polarization (particularly for small f / # lenses) and, in reality, light incident normally could cause significant RA. Broadly speaking, the reasons for this are (1) vector diffraction effects, which result in a component of p-polarization at focus [21] and (2) non-normal incidence of the peripheral rays in the light beam. For long pulses, ripples develop in the critical surface, which tend to nullify the polarization dependence of absorption [22]. This effect can be ignored in subpicosecond-laser-matter interaction, as ion-density fluctuations do not build up during the interaction time, unless a strong pre-pulse exists. However, for a macroscopically rough surface (average local roughness > λ), local depolarization of light comes in to effect. Thus, for angles close to normal incidence, large enhancements in yields can be observed by using macroscopically rough surfaces, as the local modifications of polarization facilitate RA, which was inherently absent in a polished target for s-polarized light fields or for p-polarized light at angles close to normal incidence, as shown in Fig.3 (inset). As the angle of incidence is increased, p-polarized light couples more to the plasma formed on a polished target, up to an optimum angle $\theta_{max}$ decided by the plasma length and wavelength of incident light [16]. The local depolarization adversely affects the laser absorption in a rough target, at the optimum angle deduced for the polished one. As a result of these; the yield enhancement factor reduces as the angle of incidence is increased. However, the enhancement factor does not converge to unity even at $\theta_{max}$, where one expects the absorption to be maximum in a polished target and minimum in a rough surface, due to depolarization effects. This implies that, even though the local depolarization detrimentally affects the enhancement, a rough surface still couples light better than a polished one. Thus one has to invoke extra absorption mechanisms existing only for modulated surfaces. Extra absorption mechanisms viz. surface waves and the resultant local field modifications have been shown to enhance absorption [13, 23].

Plasma equilibration is known to occur at time scales longer than the ultrashort pulse duration and therefore such a laser-produced-plasma can have electron distributions with widely different peak energies. The temperature of the electrons produced by RA is given by the scaling law [24] $T_h$ (keV) ≈ 6 × 10$^{-5}$ (I$\lambda^2$ (Wcm$^{-2}$μm$^2$))$^{0.33}$, which yields a temperature component ~ 10 keV, with our experimental parameters. The lower temperature component in the spectrum is very close to this. Recent studies [24, 25] report the existence of a bi-Maxwellian electron distribution, with widely different hot electron temperatures, similar to the spectrum in Fig. 3. The higher temperature component is believed to be that of hot electrons produced by another mechanism known as vacuum heating (VH) [26]. Recent experiments [25, 27] and simulations [25, 28] show that VH is not negligible in experimental conditions similar to ours. Though a scale length $L \ll \lambda$ is desirable for VH, Dong et al. [25] report that VH will play a crucial role in coupling the laser energy to the plasma up to $L = 0.1 \lambda$. The higher component observed in our spectrum is in close agreement with the values reported from these experiments and simulations.

Note that there is a well-defined temperature component even in the spectrum obtained from the reasonably polished surface (~ 25 nm roughness). The hot electron production in the case of the polished target at near normal incidence could be due to possible de-polarization while focusing and an induced RA due to the huge magnetic fields generated in the plasma [29, 30]. However, one expects a greater contrast (more that 4-fold enhancement) between a highly rough target and a well-polished one at $10^o$ (Fig. 3: inset). Thus one can deduce that even a feeble amount of local roughness (few tens of nanometers) affects the emission significantly. It is important to note that the roughness levels mentioned here are the local (within the focal spot) and are different from the $\lambda/n$ – polish that is normally specified for the entire surface.

We now examine the laser coupling in modulated targets at a large angle of incidence, with s-polarized light field. Fig. 4 shows the variation of bremsstrahlung emission from copper surfaces at $10^{16}$ Wcm$^{-2}$, irradiated at 50º incidence, as a function of the local surface roughness. The first curve is the emission from a reasonably polished target of average roughness ~ 25 nm and the last curve from a rough copper surface of average roughness > 1μm. The curves in between correspond to emissions from targets of intermediate roughness. The surfaces of different finishes are prepared by etching the polished surface with fine carborundum powder of different particle sizes. The local roughness scales are determined by Atomic Force Microscopy (AFM). The roughness levels are more or less uniform all over the surface.

It can be inferred from the figure that both the integrated yield as well as the hot electron temperature increase as the roughness level on the surface is increased. Under our experimental conditions, hot electron production is expected to occur for p-polarized light incident obliquely on optically polished targets, but *not for s-polarized light*. For a polished target irradiated with s-polarized light field, IB is the only major laser coupling mechanism, which is maximum at normal incidence and reduces as $1 - \exp(-\cos^5\theta)$, with the angle of incidence $\theta$, for a linear plasma density profile [16]. Hence, for a surface with a local roughness scales much less than λ, the light coupling and hence the hot electron production should be minimal, when irradiated with s-polarized light, especially at large angles of incidence. In the present study, however, we observe significant hot electron production from targets with feeble roughness levels (only 5 times larger than the optically polished ones). This again calls for the extra absorption mechanisms induced by sub-wavelength surface structures, as the depolarization by the structures would not be predominant here. As the roughness levels are increased, local depolarization facilitates RA, which will in turn enhance the hot electron production and their temperature. Thus, even with reasonable (hundreds of nanometers) roughness levels, the emission becomes polarization independent. However, as described earlier, local depolarization alone is insufficient to explain the enhancements observed as compared to polished targets and extra absorption mechanisms like surface wave coupling and local field enhancements need to be invoked. These mechanisms are well allowed by the surface roughness, irrespective of the light polarization.

From the above discussion on the role of polarization of the light field in laser-plasma coupling, one expects a drastic difference in hot electron production with s and p-polarized light fields, in an optically polished surface. While the efficiency of hard x-ray generation with p-polarized light in a polished target of local roughness ~ 5 nm is more than five times that with it's s-polarized counterpart [12], we have observed that this difference comes down significantly by increasing the local roughness levels and gets nullified even at sub-wavelength levels (hundreds of nanometers) of roughness on the surface. Further studies are expected for deeper understanding of this behaviour, but from the present measurements, one can readily infer that hot electron generation from rough surfaces is independent of the polarization of the exciting light.

## 4. Conclusions

We have investigated the hot electron production from copper surfaces of a wide range of surface finish. The polarization and the angular dependence of the emission is clearly brought out. The hot electron temperatures suggest the influence of resonance absorption (RA) and vacuum heating (VH) in the coupling of light to the plasma. Hard x-ray production with s-polarized light is studied with a surface transition from smooth to rough. Even sub-wavelength structures are shown to affect the hot-electron production from plasmas. Modulated surfaces produce enhanced yields, which are found to be independent of the incident light polarization.

## Acknowledgement

The authors acknowledge A. S. Sandhu and A. K. Dharmadhikari for valuable discussions and help in experiments. PPR acknowledges the financial support from the TIFR endowment fund. The $T^4$ system (TIFR Table Top Terawatt) has received substantial funding from the Dept. of Science and Technology, New Delhi.

**Figure Captions**

Fig.1: Schematic diagram of the experimental set-up. T – Target, P – Plasma, L – Lens, WP – Wave Plate, BS – Beam Splitter, PD – Photo Diode, HV - High Voltage, G – Gate& Delay Generator, SA – Spectroscopy Amplifier, ADC – Analog to Digital Converter, MCA – Multi Channel Analyzer. Inset: Probe reflectivity vs. time delay with pump pulse.

Fig. 2: Variation of bremsstrahlung yield with the angle of incidence from an optically polished copper target irradiated with p-polarized light. Inset: AFM image of the polished target.

Fig. 3: Bremsstrahlung emission at $10^{16}$ Wcm$^{-2}$, from a reasonably polished and rough targets using p – polarized light incident at $10^o$. Inset: The variation of the yield enhancement factor with angle of incidence using a rough target in place of an optically polished target.

Fig. 4: Bremsstrahlung emission from copper surfaces of various roughness levels, irradiated with s-polarized light at $10^{16}$ Wcm$^{-2}$.

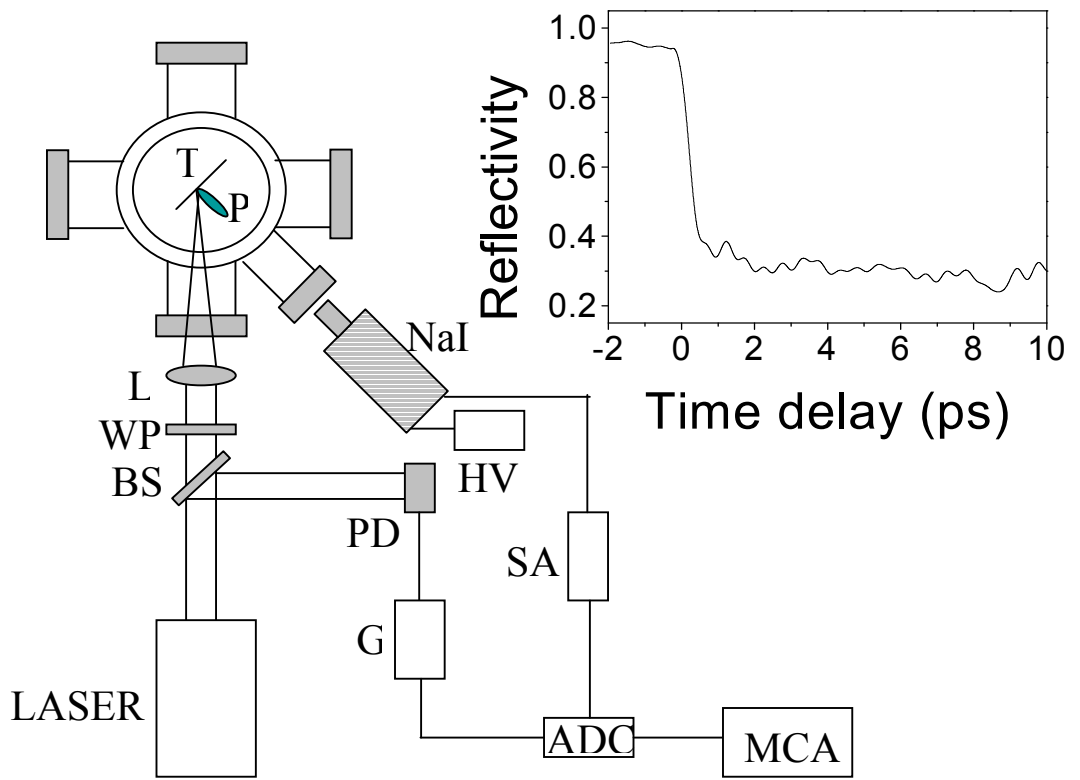

**Figure 1**

*P. P. Rajeev et al.*

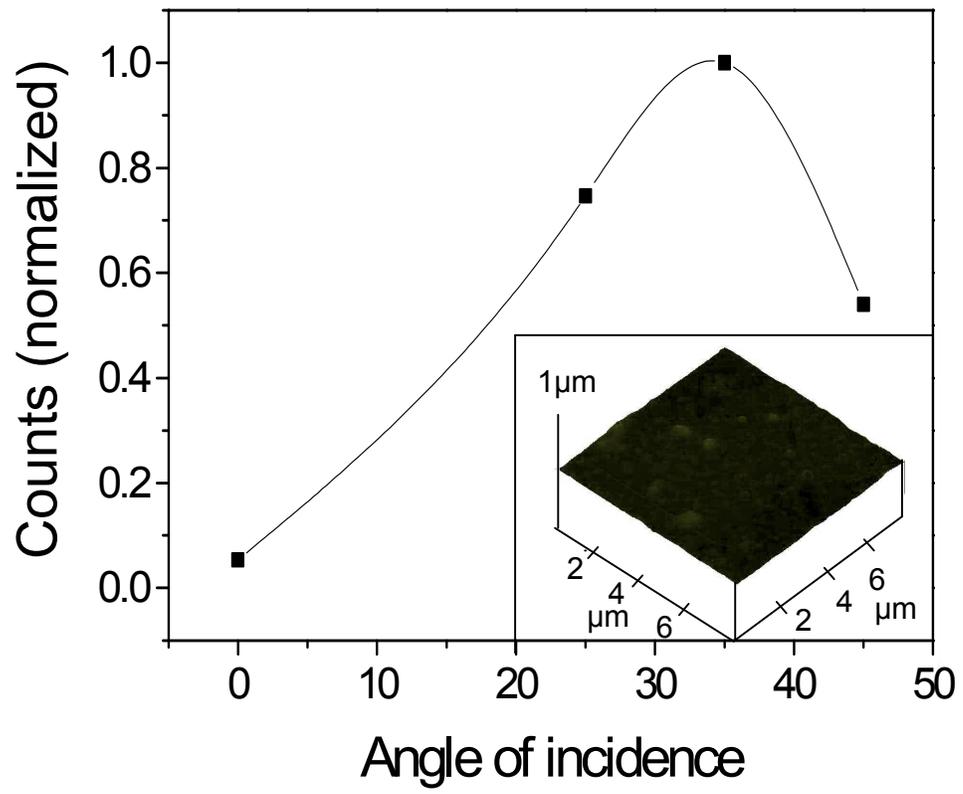

**Figure 2**

*P. P. Rajeev et al.*

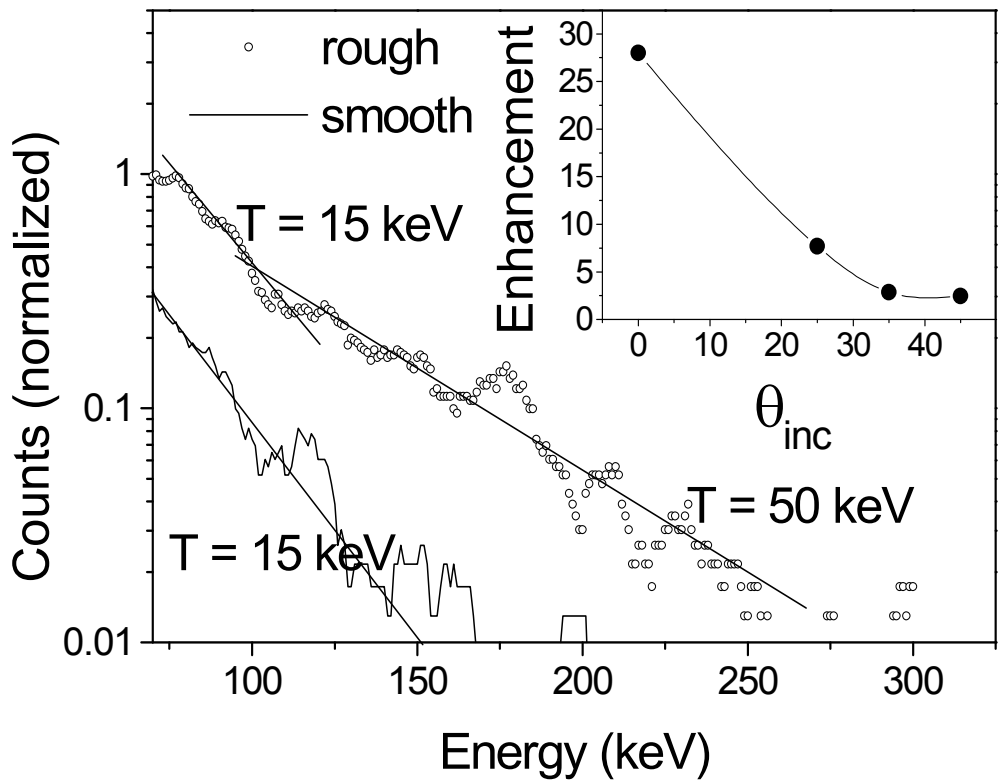

**Figure 3**

P. P. Rajeev et al.

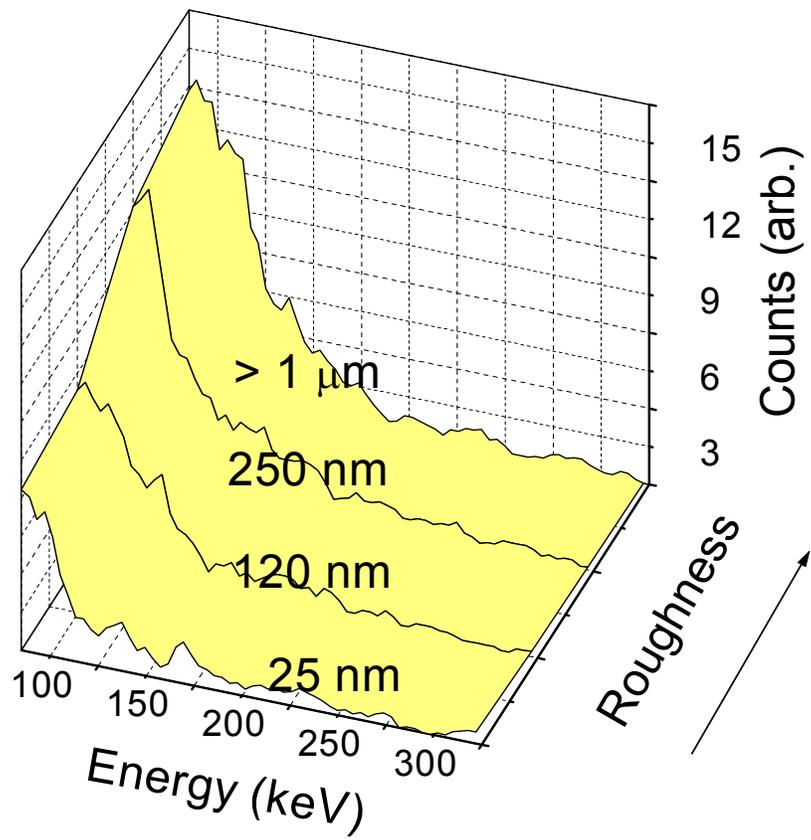

**Figure 4**

*P. P. Rajeev et al.*